\documentclass{PoS}
\rightline{\parbox{4cm}{\large\rm
ADP-14-41/T900\\
DESY 14-243\\
Edinburgh 2014/22\\
LTH 1032
}}

\usepackage{amsmath}
\usepackage{soul}
\usepackage{caption}
\usepackage{cite}

\graphicspath{{.}{graphics/}}

\title{Connected and disconnected quark contributions to hadron spin}
\ShortTitle{Hadron Spin Structure}

\author{A.~J.~Chambers$^a$,
R.~Horsley$^b$,
Y.~Nakamura$^c$,
H.~Perlt$^d$,
D.~Pleiter$^{ef}$,
P.~E.~L.~Rakow$^g$,
G.~Schierholz$^h$,
A.~Schiller$^c$,
H.~St\"uben$^i$,
R.~D.~Young$^a$
and \speaker{J.~M.~Zanotti}$^a$ \\
\llap{$^a$}CSSM, Department of Physics, University of Adelaide, Adelaide SA
5005, Australia \\
\llap{$^b$}School of Physics and Astronomy, University of Edinburgh,
Edinburgh EH9 3JZ, UK \\
\llap{$^c$}RIKEN Advanced Institute for Computational Science, Kobe, Hyogo
650-0047, Japan \\
\llap{$^d$}Institut f\"ur Theoretische Physik, Universit\"at Leipzig, 04103
Leipzig, Germany \\
\llap{$^e$}JSC, J\"ulich Research Centre, 52425 J\"ulich, Germany \\
\llap{$^f$}Institut f\"ur Theoretische Physik, Universit\"at Regensburg,
93040 Regensburg, Germany \\
\llap{$^g$}Theoretical Physics Division, Department of Mathematical Scienc
University of Liverpool, Liverpool L69 3BX, UK \\
\llap{$^h$}Deutsches Elektronen-Synchrotron DESY, 22603 Hamburg, Germany \\
\llap{$^i$}Regionales Rechenzentrum, Universit\"at Hamburg, 20146 Hamburg,
Germany \\
E-mail: \email{alexander.chambers@adelaide.edu.au},
\email{james.zanotti@adelaide.edu.au}
}

\abstract{
  By introducing an external spin operator to the fermion action, the
  quark spin fractions of hadrons are determined from the linear
  response of the hadron energies using the Feynman-Hellmann (FH)
  theorem.
  At our SU(3)-flavour symmetric point, we find that the connected
  quark spin fractions are universally in the range 55-70\% for vector
  mesons and octet and decuplet baryons.
  There is an indication that the amount of spin suppression is quite
  sensitive to the strength of SU(3) breaking.

  We also present first preliminary results applying the FH technique to
  calculations of quark-line disconnected contributions to hadronic
  matrix elements of axial and tensor operators.
  At the SU(3)-flavour symmetric point we find a small negative
  contribution to the nucleon spin from disconnected quark diagrams,
  while the corresponding tensor matrix elements are consistent with
  zero.
}

\FullConference{The 32nd International Symposium on Lattice Field Theory,\\
		23-28 June, 2014\\
		Columbia University New York, NY}

\begin{document}

\section{Introduction}

Ever since the announcement by the European Muon Collaboration (EMC)
\cite{Ashman:1987hv} that quarks carry only a small fraction of the
proton's spin, identifying the origin of hadronic spin has proven to
be a fascinating challenge (see e.g. \cite{Thomas:2009gk}).
Since this is an inherently nonperturbative phenomenon, lattice QCD
provides the ideal framework for investigating the spin decompostion
of hadrons (see \cite{Constantinou:2014tga} for a review presented at
this conference).

Standard methods for performing such a lattice QCD simulation involve
calculating hadronic three-point functions via sequential source
methods \cite{Can:2015hda}.
There has been increasing discussion surrounding the need for
controlling potential excited state contamination in these three-point
simulations, in particular for spin related quantities, such as $g_A$
\cite{Owen:2012ts,Capitani:2012gj,Dinter:2011sg,Bhattacharya:2013ehc,Bali:2013nla}.
Further challenges present themselves when calculting the
contributions from quark-line disconnected diagrams (e.g. $\Delta s$),
as these require the determination of computationally demanding
all-to-all propagators.
This problem is usually confronted through the use of stochastic
methods, and there has been a lot of recent progress in this direction
\cite{Babich:2010at,QCDSF:2011aa,Engelhardt:2012gd,Abdel-Rehim:2013wlz,Deka:2013zha}.

Recently we have proposed an alternative method for tackling these
issues through the application of the Feynman-Hellmann (FH) theorem to
lattice QCD calculations of hadronic matrix elements
\cite{Horsley:2012pz,Chambers:2014qaa}, in a similar way to that
proposed in \cite{Detmold:2004kw}.
We have also recently shown how it is possible to compute
flavour-singlet renormalisation constants nonperturbatively by an
appropriate application of the FH theorem \cite{Chambers:2014pea}.

In this talk, we present an update of the quark-line connected
contributions to the spin of various hadrons first published in
\cite{Chambers:2014qaa}.
We will also reveal first simulations of the quark-line disconnected
spin contributions through the generation of a new set of background
gauge field configurations including the axial operator in the
sea-quark action.

\section{The Feynman-Hellmann Theorem}
\label{sec:fh_theorem}
The Feynman-Hellmann theorem allows hadronic matrix elements
to be calculated from shifts in the hadron spectrum.
In general, if the action of our theory depends on some parameter $\lambda$,
then for any hadron state $H$ we have
\begin{equation}
	\frac{\partial E_H}{\partial \lambda}
	= \frac{1}{2 E_H} \left\langle H \left| \frac{\partial S}{\partial
	\lambda} \right| H \right\rangle_\lambda \, ,
	\label{eq:feynman_hellmann_theorem}
\end{equation}
where we use a subscript to indicate the $\lambda$ dependence of the matrix
element on the right-hand side of Eq.~\ref{eq:feynman_hellmann_theorem}.
In particular, if we modify the QCD action such that
\begin{equation}
	S \to S'(\lambda) = S + \lambda \int \mathrm{d}^4 x \; O(x) \, ,
	\label{eq:feynman_hellmann_modification}
\end{equation}
where $O$ is some operator, then
\begin{equation}
	\left. \frac{\partial E_H}{\partial \lambda} \right|_{\lambda=0}
	= \frac{1}{2 E_H} \left\langle H \left| O \right| H
	\right\rangle \, ,
	\label{eq:feynman_hellmann_calculation}
\end{equation}
noting the matrix element is now evaluated with respect to the unmodified
action.

The matrix element in Eq.~\ref{eq:feynman_hellmann_calculation} can be calculated
by measuring shifts in the energy of the state $H$ as the parameter
$\lambda$ is modified.
A well-known example of this approach is the calculation of
nucleon $\sigma$ terms, where the variational parameter(s) are the quark
masses (see \cite{Young:2009ps} for a review).

The Feynman-Hellmann method described can in general allow calculation of
both connected and disconnected contributions to matrix elements.
If the addition to the action in Eq.~\ref{eq:feynman_hellmann_modification}
is made during gauge field generation, one may make contact with
the disconnected contributions, while modifications to the Dirac operator
before calculating propagators allow access to the connected contributions.


\section{Lattice Details}

We use gauge field configurations with $2+1$ flavours of
non-perturbatively $O(a)$-improved Wilson fermions and a
lattice volume of $L^3 \times T = 32^3 \times 64$.
The lattice spacing $a = 0.074(2)$~fm is set using a number of singlet
quantities \cite{Horsley:2013wqa,Bietenholz:2010jr,Bietenholz:2011qq}.
The clover action used comprises the tree-level Symanzik improved
gluon action together with a stout smeared fermion action, modified
for the implementation of the Feynman-Hellmann method
\cite{Chambers:2014qaa}.
For the quark-line connected results, we use ensembles with three sets
of hopping parameters, $(\kappa_l,\kappa_s) =$
(0.120900,120900), (0.121040,120620), (0.121095,0.120512),
corresponding to pion masses in the range 470-310 MeV.

The exploratory investigation of the quark-line disconnected
contribution to the proton spin is performed at the SU(3) symmetric
point ($\kappa_l = \kappa_s = 0.120900$) where all three quarks have
the same mass ($m_\pi\approx 470$~MeV) and with two non-zero values of
$\lambda$ applied equally to all three sea quarks.

In order to presently physically relevent results, we use recent
nonperturbative determinations of the flavour non-singlet
\cite{Constantinou:2014fka} and singlet \cite{Chambers:2014pea} axial
current renormalisation constants.

\section{Connected Spin Contributions}
\label{sec:conn}

\begin{figure}
  \begin{minipage}[b]{0.42\textwidth}
	\centering
	\includegraphics[width=1.0\linewidth]{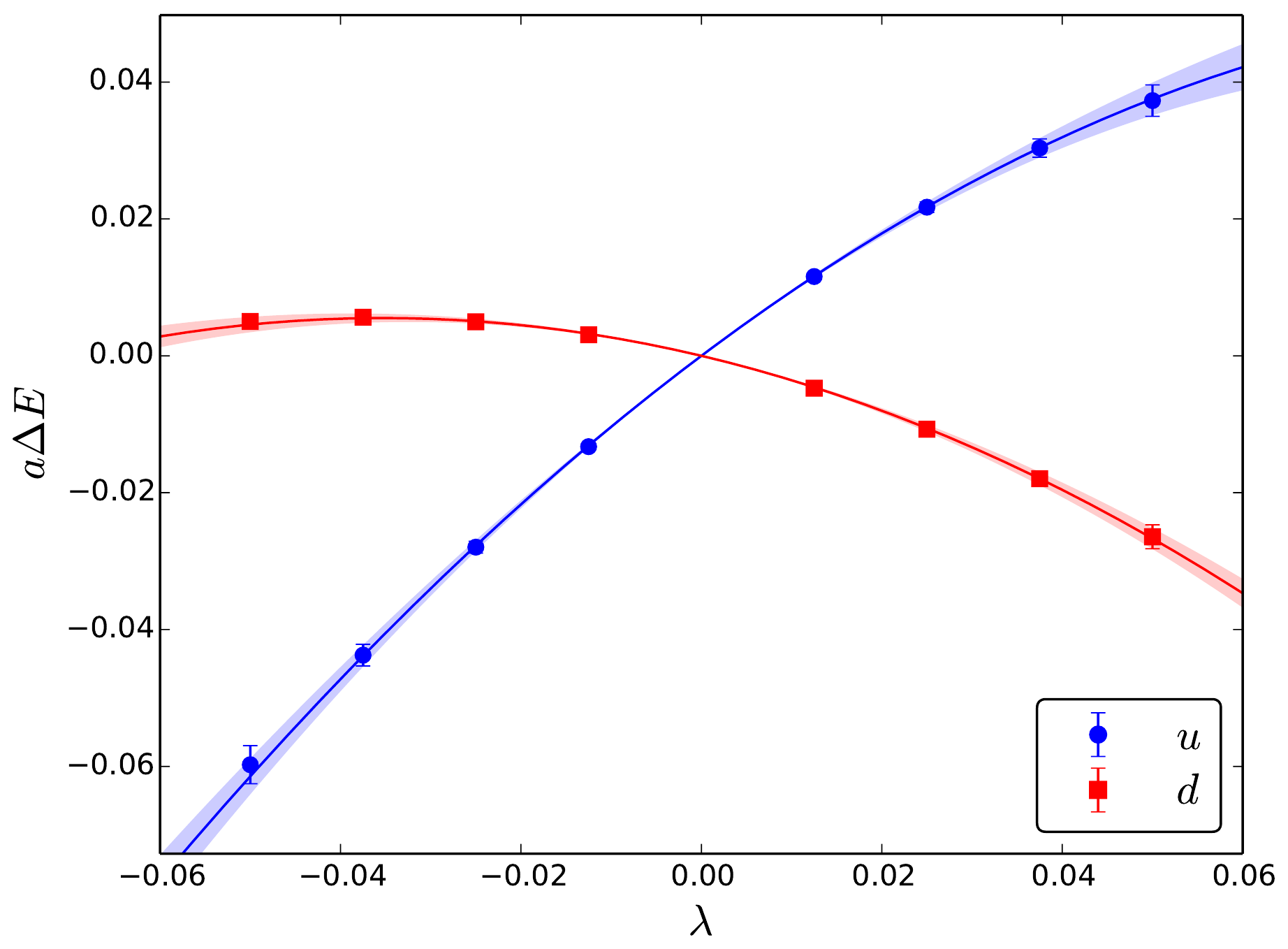}
	\captionof{figure}{Nucleon energy shifts with respect to $\lambda$ at the
	  SU(3) symmetric point.
	}
	\label{fig:nucleon_connected}
  \end{minipage}\hfill
  \begin{minipage}[b]{0.53\textwidth}
	\centering
	\includegraphics[width=1.0\linewidth]{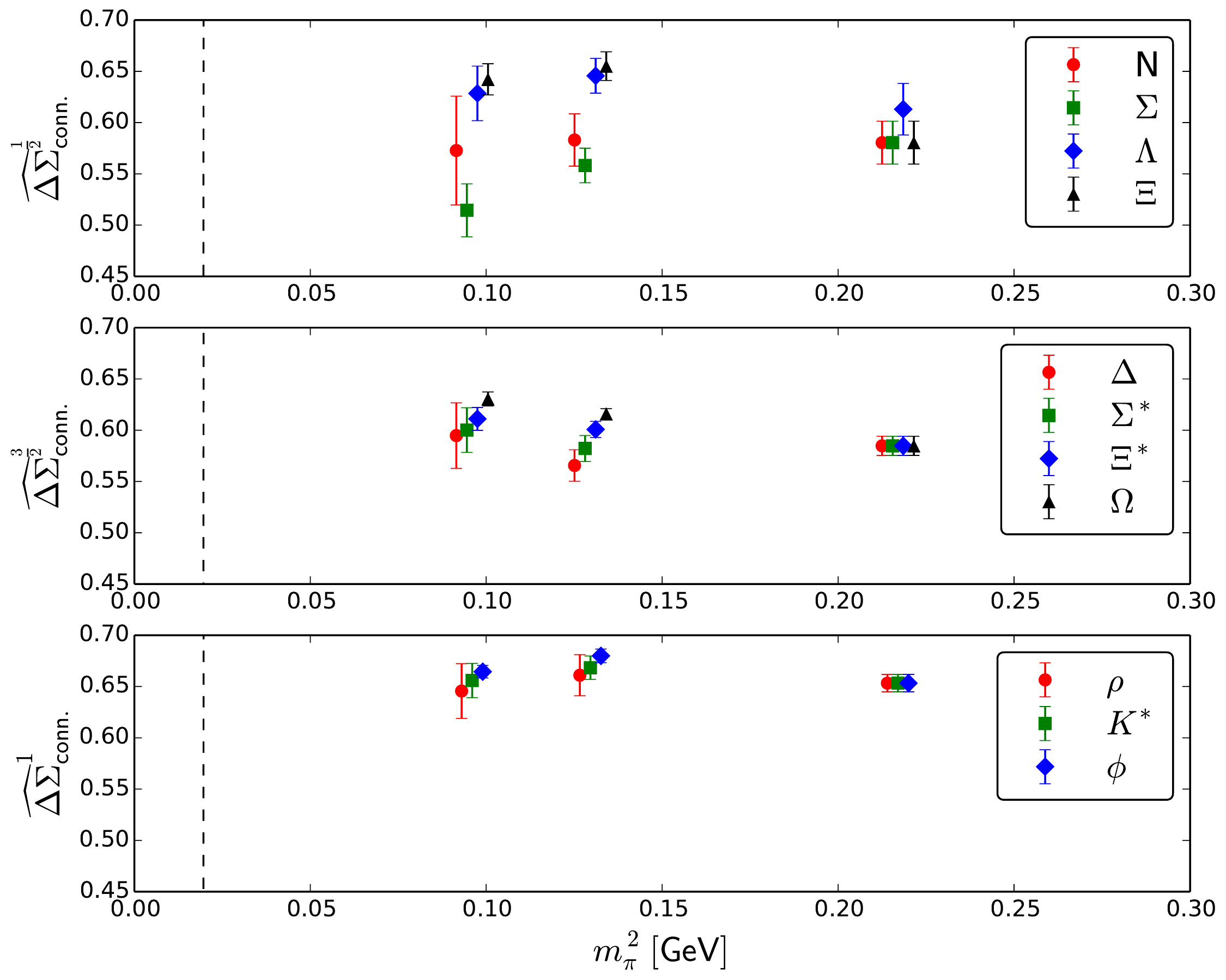}
	\captionof{figure}{Total connected quark spin contributions to various
	  hadrons as a function of pion mass.}
	\label{fig:chiral}
  \end{minipage}\hfill
\end{figure}

To calculate connected contributions to the quark axial charges using the
Feynman-Hellmann method,
the fermion matrix for a single quark flavour is modified such that
\begin{equation}
  M \to M'(\lambda) = M + \lambda i \gamma_5 \gamma_3 \, .
  \label{eq:axial_modification}
\end{equation}
where $i \gamma_5 \gamma_3$ is the Euclidean form of the axial operator.
Hence, for a general zero-momentum hadron state $H$,
polarized in the $z$-direction, we have
\begin{equation}
  \left. \frac{\partial E_H}{\partial \lambda} \right|_{\lambda=0}
  = \frac{1}{2 M_H} \left\langle H \, ; \, J \, m
  \left|\bar{q} i \gamma_5 \gamma_3 q \right|
  H\, ; \, J \, m \right\rangle
  = \Delta q^{Jm} \, .
  \label{eq:axial_calculation}
\end{equation}
Here $J$ and $m$ are the spin and longitudinal spin polarisation quantum
numbers, respectively. See \cite{Chambers:2014qaa} for a full discussion of this
notation.
Since the gauge fields used in this simulation were not generated with the modified
operator in Eq.~\ref{eq:axial_modification}, we do not access disconnected
contributions to the axial charges (as discussed in Sec.~\ref{sec:fh_theorem}).
We also note from Eq.~\ref{eq:axial_modification} that reversing the spin of the
hadron state is equivalent to flipping the sign of $\lambda$. So we may easily
double our sampled parameter space by identifying, for example, spin-up nucleon states with
positive $\lambda$, and spin-down states with negative $\lambda$.
We improve the extracted signals by forming ratios of correlation functions at
$\lambda \ne 0$ and $\lambda = 0$, the details of which are discussed in
\cite{Chambers:2014qaa}.
Fig.~\ref{fig:nucleon_connected} shows nucleon energy shifts for different
values of $\lambda$, in the cases where the modification to the fermion matrix in
Eq.~\ref{eq:axial_modification} is made to the $u$ or $d$ part separately.

An important advantage of the Feynman-Hellmann method is that matrix elements of
a particular operator may be easily calculated for different hadrons, without
additional inversions being required. Fig.~\ref{fig:chiral} shows results at
three
different pion masses for the total connected quark axial charges in various
different hadrons.

Further discussion of these calculations may be found in \cite{Chambers:2014qaa}.

\section{Disconnected Spin}

In principle, the application of the Feynman-Hellmann method to disconnected
contributions is no less straightforward than the above, requiring only the
modification of the fermion action during gauge field generation.
However, an issue arises because the axial operator does not satisfy
$\gamma_5$-hermiticity, and hence the modification to the Dirac operator in
Eq.~\ref{eq:axial_modification} generates a sign problem.
To avoid this, we instead modify the fermion matrix such that
\begin{equation}
  M \to M'(\lambda) = M + \lambda \gamma_5 \gamma_3 \, .
\end{equation}
With this modification, the signal manifests as a complex phase in the
correlation function
\begin{equation}
   C(\lambda,t) = A e^{-E t} e^{i\phi t} \, ,
   \label{eq:disconn_corr}
\end{equation}
where any shift in $\phi$ with respect to $\lambda$ is related to the axial
charge by
\begin{equation}
  \left. \frac{\partial \phi}{\partial \lambda} \right|_{\lambda=0} = \Delta
  q^{Jm} \, .
  \label{eq:disconn_relation}
\end{equation}
To first order in $\lambda$, the coefficient $A$ in Eq.~\ref{eq:disconn_corr} is real and there is no shift
in $E$. This motivates the ratio
\begin{equation}
  R(\lambda,t) = \frac{\operatorname{Im}C(\lambda,t) -
  \operatorname{Im}C(-\lambda,t)}{\operatorname{Re}C(\lambda,t) +
  \operatorname{Re}C(-\lambda,t)}
  = \tan \phi t \, ,
  \label{eq:disconn_ratio}
\end{equation}
where $C(-\lambda,t)$ is the spin-down hadron state at positive $\lambda$,
recalling the discussion in Sec.~\ref{sec:conn}.

\begin{figure}
	\begin{minipage}[t]{0.42\textwidth}
	\centering
	\includegraphics[width=1.0\linewidth]{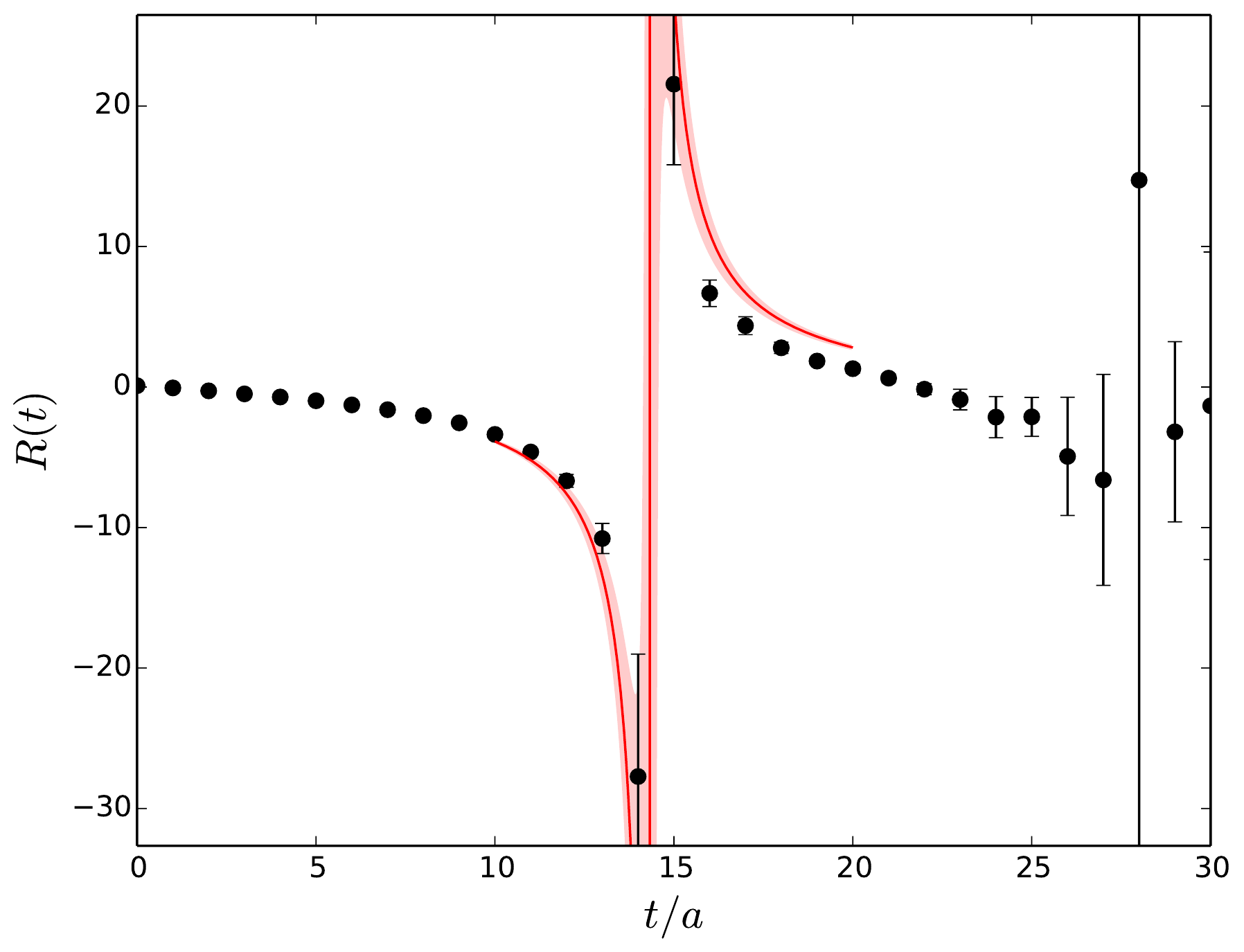}
	\captionof{figure}{$R(\lambda,t)$ for the nucleon as defined in
	  Eq.~\protect\ref{eq:disconn_ratio} for $\lambda_{\text{conn.}}=-0.1$, at the SU(3)
	  symmetric point.}
	\label{fig:tan}
  \end{minipage}\hfill
  \begin{minipage}[t]{0.53\textwidth}
	\centering
	\includegraphics[width=1.0\linewidth]{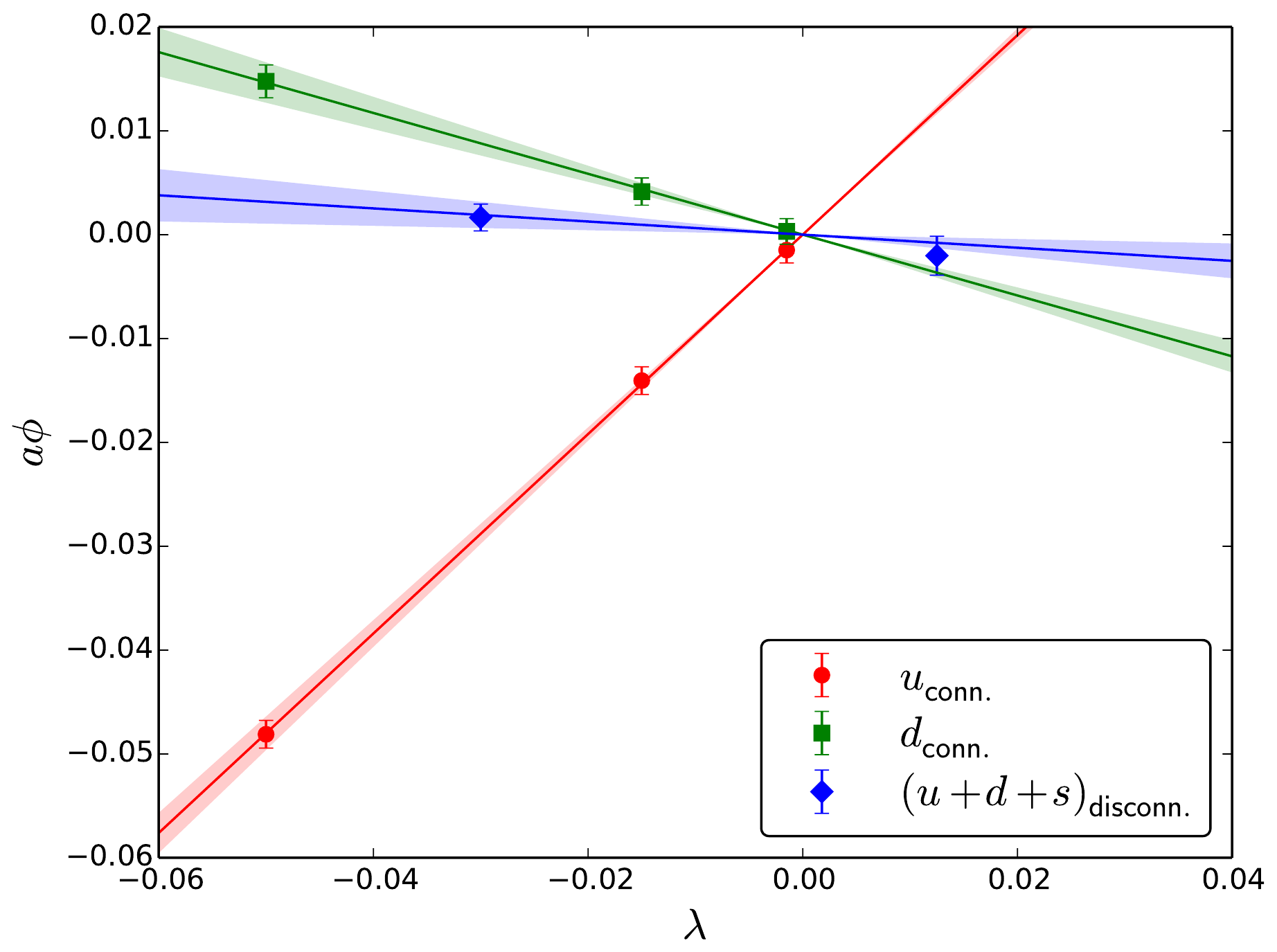}
	\captionof{figure}{Complex phase of the nucleon correlator as a function of $\lambda$
	  at the SU(3) symmetric point. Connected and disconnected calculations
	are shown.}
	\label{fig:disconn}
  \end{minipage}\hfill
\end{figure}

A plot of this ratio for a large value of $\lambda=-0.1$ can be seen in
Fig.~\ref{fig:tan}.
With this background field strength, it is possible that terms beyond linear
order in $\lambda$ influence the ratio, which would deviate from
pure tangential behaviour. Nevertheless, we still observe the general tangent
form.
At the smaller $\lambda$ used for the Feynman-Hellmann calculation, this
higher-order behaviour is not an issue.

Fig.~\ref{fig:disconn} shows the change in the complex phase extracted
from the tangential fit, with the
background field strength. We are able to reproduce results for
the connected contributions to the axial charges on the same ensemble (from
\cite{Chambers:2014qaa}, shown in brackets),
\begin{align}
  \Delta u_{\text{conn.}}(m_{\pi} \approx 470 \text{ MeV})
  & = \phantom{-}0.816(33) \, ,
  \qquad \qquad & [\phantom{-}0.849(17)] \\
  \Delta d_{\text{conn.}}(m_{\pi} \approx 470 \text{ MeV})
  & = -0.249(34) \, .
  & [-0.268(12)]
\end{align}
Using configurations generated with non-zero $\lambda$ applied to the sea
quarks, we calculate from a linear fit to the blue data in
Fig.~\ref{fig:disconn}, the total disconnected quark contribution to the
axial charges,
\begin{equation}
	\Delta \Sigma_{\text{disconn.}}(m_{\pi} \approx 470 \text{ MeV}) =
	-0.055(36) .
\end{equation}
Here we make use of the singlet axial current renormalisation constant
calculated in \cite{Chambers:2014pea}. This is in agreement with
stochastic estimations of this value reported in \cite{QCDSF:2011aa}.

\section{Tensor Charge}

Calculation of the quark contributions to the tensor charge are calculated
with the previously described methods. The fermion matrix is
modified such that
\begin{equation}
  M \to M'(\lambda) = M + \lambda \gamma_5 \sigma_{34}\, ,
  \label{eq:tensor_mod}
\end{equation}
so that the corresponding energy shifts are given by
\begin{equation}
  \left. \frac{\partial E_H}{\partial \lambda} \right|_{\lambda=0}
  = \frac{1}{2 M_H} \left\langle H \, ; \, J \, m \left|\bar{q} \gamma_5
  \sigma_{34} q
  \right| H\, ; \, J \, m \right\rangle
\end{equation}
We note that since the tensor operator flips helicity, the disconnected
insertions of the operator must vanish in the chiral limit. However, away
from the chiral limit, this is no longer guaranteed.

\begin{figure}
	\centering
	\includegraphics[width=0.6\linewidth]{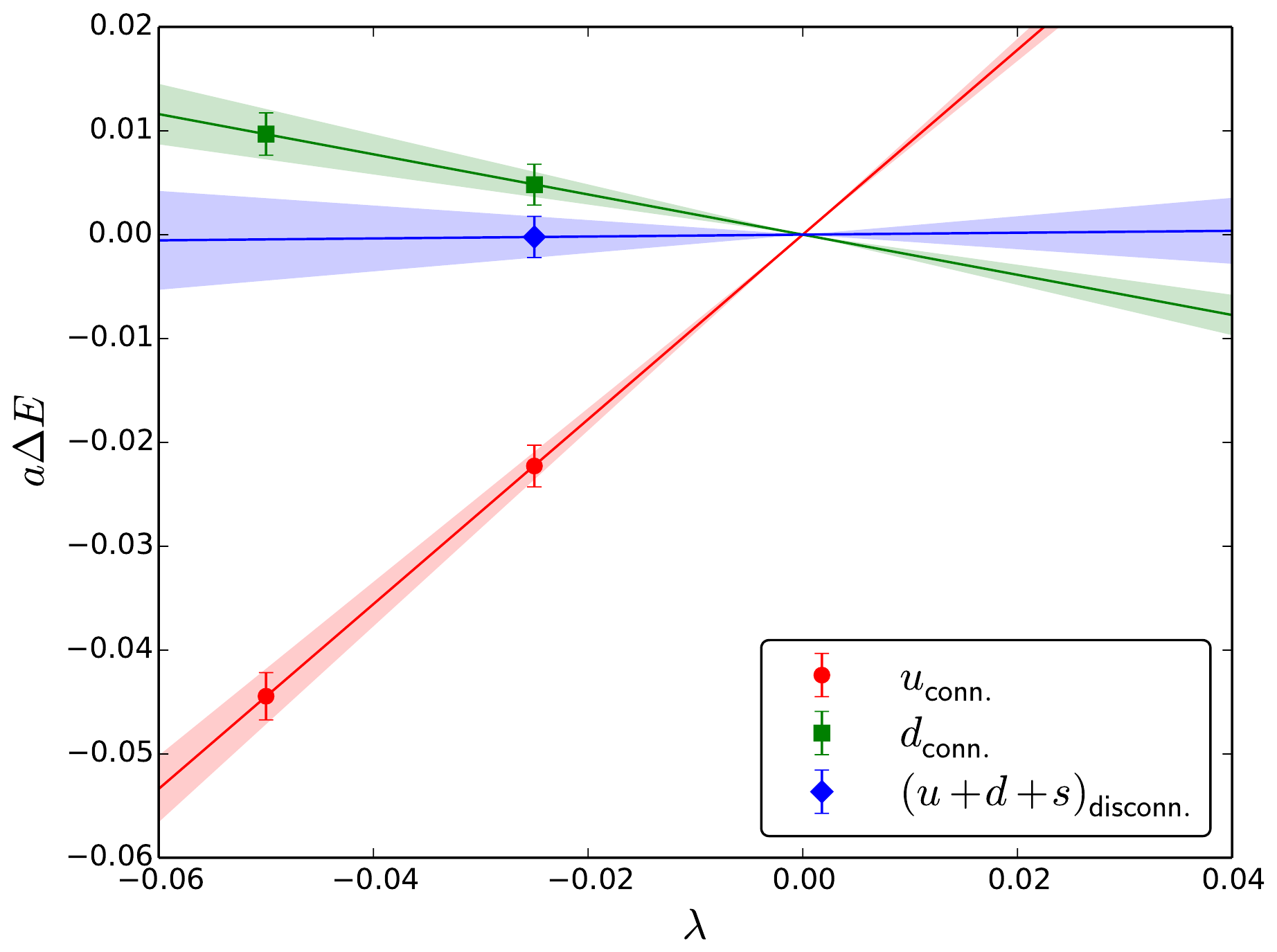}
	\captionof{figure}{Nucleon energy shift with $\lambda$ when the
	  modification described in Eq.~\protect\ref{eq:tensor_mod} is made.}
	\label{fig:tensor}
\end{figure}

Fig.~\ref{fig:tensor} shows nucleon energy shifts for the cases where the
fermion matrix is modified before inversion (accessing the connected
contributions to the tensor charge), and in the case where modified gauge
fields are used to access the disconnected contributions.
From a linear fit to the data, we calculate connected contributions
\begin{align}
	\delta u_{\text{conn.}}(m_{\pi} \approx 470 \text{ MeV}) & = \phantom{-}0.881(17) \, ,\\
	\delta d_{\text{conn.}}(m_{\pi} \approx 470 \text{ MeV}) & = -0.198(12) \, ,
\end{align}
and disconnected contributions
\begin{equation}
	\delta (u + d + s)_{\text{disconn.}}(m_{\pi} \approx 470
        \text{ MeV}) = -0.009(79) ,
\end{equation}
which is consistent with zero.
\acknowledgments

The numerical configuration generation was performed using the BQCD
lattice QCD program, \cite{Nakamura:2010qh}, on the IBM BlueGeneQ
using DIRAC 2 resources (EPCC, Edinburgh, UK), the BlueGene P and Q at
NIC (J\"ulich, Germany) and the Cray XC30 at HLRN
(Berlin--Hannover, Germany).
Some of the simulations were undertaken using resources awarded at the
NCI National Facility in Canberra, Australia, and the iVEC facilities
at the Pawsey Supercomputing Centre. These resources are provided
through the National Computational Merit Allocation Scheme and the
University of Adelaide Partner Share supported by the Australian
Government.
The BlueGene codes were optimised using Bagel \cite{Boyle:2009vp}.
The Chroma software library \cite{Edwards:2004sx}, was used in the
data analysis.
This investigation has been supported partly by the EU grants 283286
(HadronPhysics3), 227431 (Hadron Physics2) and 238353 (ITN
STRONGnet), and by the Australian Research Council under grants
FT120100821, FT100100005 and DP140103067 (RDY and JMZ).

\end{document}